\documentclass[conference]{IEEEtran}
\IEEEoverridecommandlockouts
\usepackage{amsmath,amssymb,amsfonts}
\usepackage[noend]{algpseudocode}
\usepackage{subcaption}
\usepackage{graphicx}
\usepackage{algorithm}
\usepackage{algpseudocode}

\usepackage{physics}

\usepackage{braket}
\usepackage{mathtools}
\usepackage{bm}
\usepackage{xcolor}
\usepackage{tikz}
\usepackage{multirow}
\usepackage{hyperref}
\usepackage{multirow}
\usepackage{array}
\usepackage{graphicx}     % 用于插入图片
\usepackage{subcaption}   % 支持 subfigure 的 caption 和布局
\usepackage{colortbl}
\usepackage{siunitx}
\usepackage{caption}

\usepackage{booktabs}

\definecolor{headercolor}{RGB}{79, 70, 229}
\definecolor{darkblue}{RGB}{30, 64, 175}

\newcommand{\correspondingauthor}{\textsuperscript{\textdagger}} % 等同贡献用匕首符号
\newcommand{\equalcontrib}{\textsuperscript{\textasteriskcentered}} % 通讯作者用星号

\def\BibTeX{{\rm B\kern-.05em{\sc i\kern-.025em b}\kern-.08em
    T\kern-.1667em\lower.7ex\hbox{E}\kern-.125emX}}
\begin{document}

\title{MolQAE: Quantum Autoencoder for Molecular Representation Learning
}

% \author{\IEEEauthorblockN{Anonymous Authors}}

\author{
\IEEEauthorblockN{Yi Pan\textsuperscript{1,*}, 
Hanqi Jiang\textsuperscript{1,*},
Wei Ruan\textsuperscript{1},
Dajiang Zhu\textsuperscript{2},
Xiang Li\textsuperscript{3},
Yohannes Abate\textsuperscript{4}, \\
Yingfeng Wang\textsuperscript{5,†}, 
Tianming Liu\textsuperscript{1,†}}
\IEEEauthorblockA{\textsuperscript{1}School of Computing, The University of Georgia, Athens, GA, USA}
\IEEEauthorblockA{\textsuperscript{2}Department of Computer Science and Engineering, The University of Texas at Arlington, Arlington, TX, USA}
\IEEEauthorblockA{\textsuperscript{3}Department of Radiology, Massachusetts General Hospital and Harvard Medical School, Boston, MA, USA}
\IEEEauthorblockA{\textsuperscript{4}Department of Physics and Astronomy, The University of Georgia, Athens, GA, USA}
\IEEEauthorblockA{\textsuperscript{5}Department of Computer Science and Engineering, University of Tennessee at Chattanooga, Chattanooga, TN, USA}

% 脚注说明
\thanks{\equalcontrib These authors contributed equally to this work.}
\thanks{\correspondingauthor Corresponding authors: Yingfeng Wang (yingfeng-wang@utc.edu) and Tianming Liu (tliu@uga.edu).}
}

% \author{
% \IEEEauthorblockN{Jingqing Ruan\IEEEauthorrefmark{1}\IEEEauthorrefmark{2}\IEEEauthorrefmark{4}, Runpeng Xie\IEEEauthorrefmark{1}\IEEEauthorrefmark{3}\IEEEauthorrefmark{4}, Xuantang Xiong\IEEEauthorrefmark{1}\IEEEauthorrefmark{3}, Shuang Xu\IEEEauthorrefmark{1}, Bo Xu\IEEEauthorrefmark{1}\IEEEauthorrefmark{2}\IEEEauthorrefmark{3}}

% \IEEEauthorblockA{\IEEEauthorrefmark{1}Institute of Automation, Chinese Academy of Sciences}
% \IEEEauthorblockA{\IEEEauthorrefmark{2}School of Future Technology, University of Chinese Academy of Sciences}
% \IEEEauthorblockA{\IEEEauthorrefmark{3}School of Artificial Intelligence, University of Chinese Academy of Sciences}
% \IEEEauthorblockA{\IEEEauthorrefmark{4}These authors contributed equally to this work.} % Custom note
% }

\maketitle

\begin{abstract}
% SMILES strings represent rich molecular structural information and quantum embedding has proven effective for processing complex information. 
We introduce Quantum Molecular Autoencoder (MolQAE), the first quantum autoencoder to leverage the complete molecular structures. MolQAE uniquely maps SMILES strings directly to quantum states using parameterized rotation gates, preserving vital structural information. Its quantum encoder-decoder framework enables latent space compression and reconstruction. A dual-objective strategy optimizes fidelity and minimizes trash state deviation.  Our evaluations demonstrate effective capture of molecular characteristics and a remarkable preservation of fidelity, approaching robust molecular reconstruction even with substantial dimensionality reduction. Our model establishes a quantum pathway in cheminformatics by being the first to process complete molecular structural information with a dedicated quantum architecture considering the Noisy Intermediate-Scale Quantum (NISQ)-era development and promising significant advances in drug and materials discovery.
\end{abstract}

\begin{IEEEkeywords}
Quantum machine learning, molecular representation learning, quantum autoencoder.
\end{IEEEkeywords}

\section{Introduction}

Molecular representation learning has emerged as a fundamental challenge in computational chemistry and drug discovery, requiring methods that can encode complex molecular structures into compact, information-rich representations suitable for downstream machine learning tasks \cite{Gilmer2017}. While classical approaches based on handcrafted descriptors and graph neural networks have achieved remarkable success \cite{Yang2019}, they face inherent limitations when processing the exponentially large conformational and electronic state spaces that characterize molecular systems \cite{Wu2018}. These limitations become particularly acute when attempting to capture quantum mechanical properties such as electron correlation and delocalization, which are crucial for accurate prediction of molecular behavior yet difficult to represent in classical feature spaces.

The fundamental quantum nature of molecular systems suggests that quantum computing may offer natural advantages for molecular representation learning \cite{Cao2018}. Quantum computers can efficiently represent superposition states and entanglement patterns that directly correspond to molecular electronic structures, potentially capturing chemical information that classical representations struggle to encode. Recent advances in quantum machine learning have demonstrated promising results in classification \cite{Havlicek2019}, regression \cite{Schuld2019}, and generative modeling \cite{Benedetti2019}, with quantum embeddings showing particular promise in capturing complex structural features that elude classical approaches \cite{PhysRevA.111.022431, PhysRevA.107.042615}. However, the application of quantum computing to molecular representation learning remains largely unexplored, despite the natural alignment between quantum mechanical properties of molecules and the computational capabilities of quantum systems.

Quantum autoencoders (QAEs) present a particularly compelling approach for molecular representation learning \cite{romero2017quantum, ma2024quantum}. Originally developed for quantum data compression, QAEs utilize parameterized quantum circuits to encode input states into lower-dimensional latent representations while preserving essential information. Unlike classical autoencoders that operate on vector representations in Euclidean space, QAEs manipulate quantum states directly in Hilbert space, enabling them to leverage quantum phenomena such as superposition and entanglement for more efficient information encoding. This quantum mechanical foundation makes QAEs especially suitable for molecular systems, where electronic wavefunctions and molecular orbitals naturally exist as quantum states.

Despite these theoretical advantages, significant challenges have prevented the practical application of quantum methods to molecular representation learning. The primary obstacle lies in effectively bridging classical molecular descriptions—such as SMILES strings or molecular graphs—with quantum state representations \cite{Kandala2017}. Previous attempts have typically relied on simplistic encoding schemes that discard crucial structural information, such as atomic connectivity and bond type. Additionally, training quantum circuits presents unique difficulties including measurement noise, limited circuit depth on near-term devices, and the presence of barren plateaus in the parameter landscape that can trap optimization in suboptimal regions \cite{McClean2018}. These challenges necessitate novel approaches that can preserve molecular information through the classical-to-quantum encoding process while maintaining trainability on current quantum hardware.

In this work, we present the Quantum Molecular Autoencoder (MolQAE), a framework that addresses these challenges through a carefully designed quantum circuit architecture for molecular representation learning. Our approach introduces three key innovations that distinguish it from previous quantum molecular methods. First, we develop a direct quantum encoding scheme that maps SMILES token sequences to parameterized quantum states, preserving sequential and structural information that frequency-based or amplitude encoding methods typically discard. Each token's positional and chemical information is encoded through quantum gate parameters, enabling the representation of complex molecular patterns including rings, branches, and stereochemistry. Second, we design a hierarchical encoder-decoder architecture with specialized processing for latent and ancillary qubits, allowing flexible compression ratios while maintaining high reconstruction fidelity. Third, we introduce a dual-objective training strategy that balances reconstruction accuracy with effective ancillary state compression, addressing the unique optimization challenges of quantum circuits through hybrid quantum-classical methods.

% We evaluate MolQAE on the QM9 dataset \cite{blum,rupp}, a standard benchmark containing 134,000 small organic molecules with up to nine heavy atoms. Our experiments demonstrate that MolQAE can compress 8-qubit molecular representations into latent spaces as small as 2-3 qubits while maintaining reconstruction fidelities exceeding 95\%. Analysis of the learned representations reveals that the quantum latent space preserves molecular similarity relationships, with chemically similar molecules clustering in the compressed representation. These results suggest that quantum autoencoders can effectively capture and compress molecular information, offering a promising direction for quantum-enhanced molecular analysis as quantum hardware continues to mature.

The \textbf{key contributions} of this paper are:

\begin{itemize}
    \item We introduce the first parameterized quantum state preparation method that directly maps SMILES token sequences to quantum states, preserving structural and sequential information critical for molecular characterization.
    
    \item We design the first quantum autoencoder circuit with specialized latent and ancillary qubit processing, enabling flexible compression ratios while maintaining quantum coherence properties essential for molecular representation.
    
    \item We develop a training approach that combines quantum circuit execution with classical parameter optimization, addressing barren plateaus and measurement noise through careful loss function design and gradient estimation techniques.
    
    \item We demonstrate through systematic experiments that quantum autoencoders can achieve significant dimensionality reduction while preserving molecular properties, establishing a foundation for future quantum molecular applications.
\end{itemize}

This work represents an important step toward practical quantum advantage in molecular sciences. By demonstrating that quantum circuits can effectively encode and compress molecular information, we establish a foundation for future applications in drug discovery, materials design, and chemical reaction prediction. As quantum hardware capabilities expand and circuit depths increase, frameworks like MolQAE may enable the exploration of larger molecular systems and more complex chemical phenomena that remain intractable for classical methods.

\section{Related Work}
% Molecular representation learning is foundational to computational chemistry, evolving from engineered features like SMILES strings towards data-driven approaches using Graph Neural Networks and autoencoders~\cite{atz2021geometric,tarahomi2025machine}. Variational Autoencoders (VAEs), in particular, excel at learning compressed, meaningful latent representations from SMILES or graphs, which benefit dimensionality reduction, downstream predictions, and generative molecular design, often enhanced via property co-learning~\cite{alperstein2023all,lin2023movae}. As a quantum analogue, Quantum Autoencoders (QAEs) aim to compress molecular information into quantum states, potentially leveraging quantum phenomena. Key challenges include encoding classical molecular data (SMILES, graphs, 3D coordinates) into suitable quantum states and defining compatible latent space distributions, like the hyperspherical von Mises-Fisher, potentially better suited to quantum state normalization than traditional Gaussian priors~\cite{zhang2024molecular,baiardi2023quantum}. While recent QAE explorations have tackled representations derived from parts of molecular information, such as molecular graphs (SQ-VAE) \cite{pmlr-v162-takida22a} and 3D atomic coordinates (QVAE-Mole) \cite{wu2024qvae}, demonstrating the emerging potential and ongoing development of quantum approaches in this domain, a quantum autoencoder processing entire molecular structures directly from sequential representations like SMILES has not yet been introduced.

\subsection{Molecular Representation Learning}

Traditional molecular representation learning methods, such as graph neural networks (GNNs) or SMILES-based models, have achieved strong performance in molecular property prediction by encoding structural features of molecules \cite{atz2021geometric,boulougouri2024molecular}. However, these approaches often neglect important quantum-level properties and suffer from limitations like poor generalization in out-of-distribution scenarios \cite{zhuang2023learning,zhang2024mvmrl}. For example, current GNNs struggle to incorporate multiscale chemical features or dynamic molecular conformations, and typically depend on large annotated datasets \cite{chen2023atomic,guo2022graph}.

\subsection{Quantum Embedding}

Quantum embedding has emerged as a powerful framework for representing classical data in quantum-enhanced feature spaces. Recent studies have validated its effectiveness in real-world scenarios. Albrecht et al. \cite{PhysRevA.107.042615} demonstrated that quantum feature maps implemented on neutral atom quantum processors can distinguish graph structures and perform molecular classification tasks with accuracy comparable to classical kernels. Ghosh et al. \cite{PhysRevA.111.022431} proposed a quantum feature mapping strategy based on the dynamics of the XY spin model, which enables effective quantum reservoir computation with enhanced expressive capacity.

\subsection{Quantum Autoencoder}

Quantum autoencoders have shown promise in compressing quantum information and dimensionality reduction for molecular dynamics \cite{romero2017quantum,cao2021noise}. While recent QAE explorations have tackled representations derived from parts of molecular information, such as 3D atomic coordinates (QVAE-Mole) \cite{wu2024qvae}, demonstrating the emerging potential and ongoing development of quantum approaches in this domain, a quantum autoencoder processing complete molecular structures directly from sequential representations like SMILES has not yet been introduced.

\section{Methodology}

\subsection{Theoretical Foundation of Quantum Autoencoders}

The quantum autoencoder (QAE) represents a fundamental paradigm shift from classical dimensionality reduction, operating directly in quantum Hilbert space to exploit quantum mechanical phenomena for enhanced molecular representation learning. Unlike classical autoencoders constrained to Euclidean vector spaces, quantum autoencoders manipulate quantum states through unitary transformations, leveraging superposition, entanglement, and quantum parallelism to achieve exponential representational advantages for molecular systems.

Given an ensemble of input quantum states $\{\rho_i\}_{i=1}^n$ with dimension $2^{N}$, a quantum autoencoder compresses these states into a latent space of dimension $2^{N_{\text{latent}}}$ where $N_{\text{latent}} < N$ through a parameterized unitary transformation $\mathcal{E}_\theta$. The compression operation can be formally expressed as $\mathcal{E}_\theta: \mathcal{H}_{\text{input}} \rightarrow \mathcal{H}_{\text{latent}} \otimes \mathcal{H}_{\text{ancilla}}$, where $\mathcal{H}_{\text{input}}$ represents the input Hilbert space of dimension $2^N$, $\mathcal{H}_{\text{latent}}$ denotes the latent space of dimension $2^{N_{\text{latent}}}$, and $\mathcal{H}_{\text{ancilla}}$ corresponds to the ancillary space of dimension $2^{N_{\text{ancilla}}}$ with $N_{\text{ancilla}} = N - N_{\text{latent}}$.

The key insight in quantum autoencoding lies in the successful disentanglement of the latent and ancillary subsystems. For optimal compression, the ancillary qubits should be driven to a fixed state $\ket{0}_{\text{ancilla}}$ independent of the input state, formally expressed as $U_\theta\ket{\psi}_{\text{input}} = \ket{\phi}_{\text{latent}} \otimes \ket{0}_{\text{ancilla}}$. This fixed state of the ancillary qubits is often referred to as the "trash state", as it ideally contains no useful information about the input state. When this condition is satisfied, all relevant information from the input state is successfully compressed into the latent representation $\ket{\phi}_{\text{latent}}$. The quantum mechanical nature of this process offers significant advantages over classical autoencoders through quantum superposition enabling exponentially compact representations, entanglement capturing complex molecular correlations, and unitary evolution ensuring information preservation within the fundamental limits of quantum mechanics.

Our proposed Molecular Quantum Autoencoder (MolQAE) leverages these quantum properties to process complete molecular structures. We instantiate this theoretical framework with $N = 8$ qubits for molecular representation, supporting variable latent dimensions $N_{\text{latent}} \in \{1, 2, ..., 7\}$ to accommodate different compression requirements. The following sections detail our implementation, beginning with the quantum encoding of molecular structures and proceeding through the encoder-decoder architecture and optimization strategy.

\begin{figure*}[htbp]
    \centering
    \includegraphics[width=0.8\textwidth]{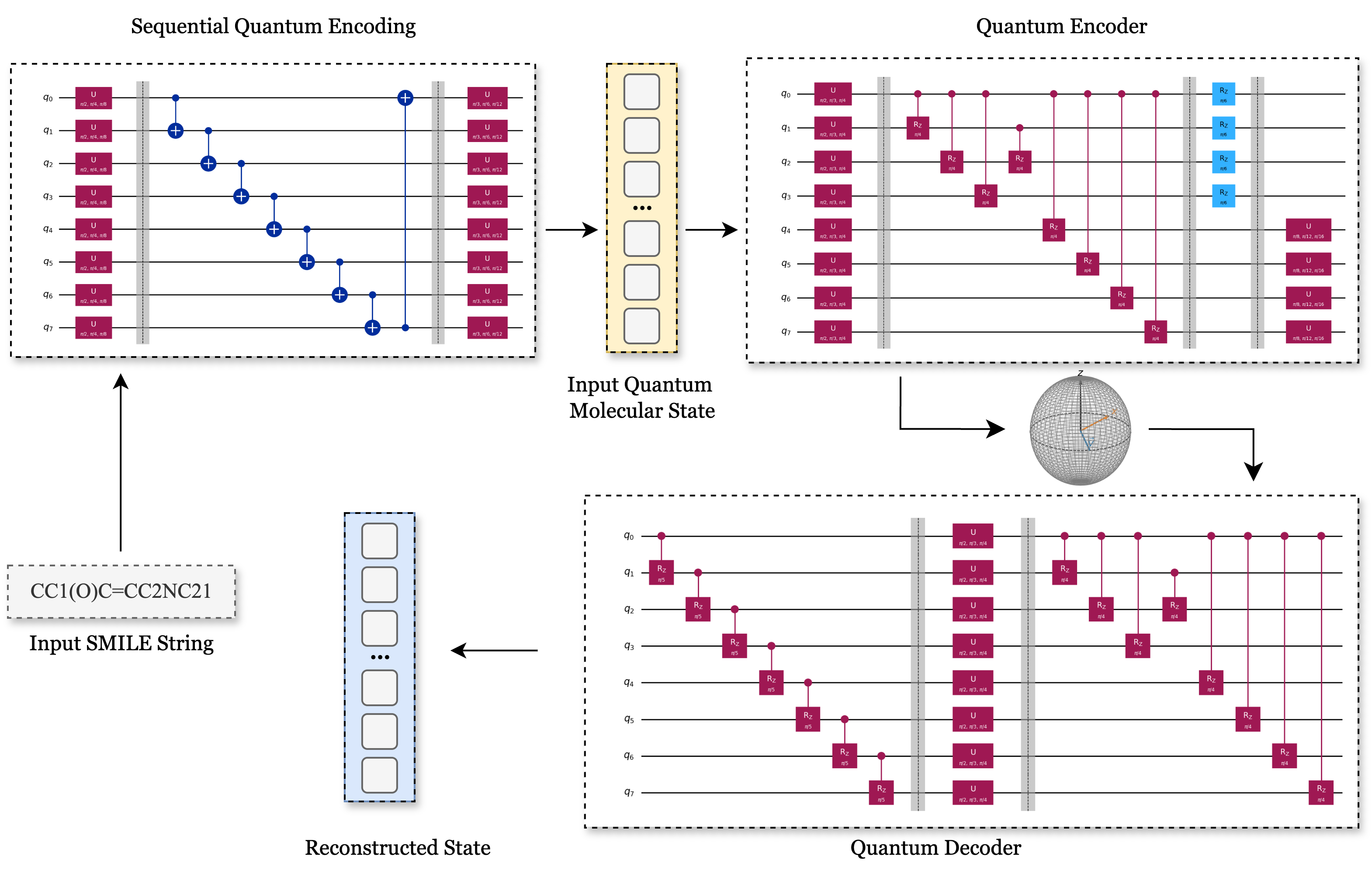}
    \caption{Framework of MolQAE illustrating the end-to-end quantum molecular compression pipeline. The system transforms SMILES strings through (a) tokenization and feature extraction, (b) quantum state preparation using parameterized $U_3$ gates and ring-topology CNOT gates, (c) multi-layer encoder compression mapping to $N_{\text{latent}}$ qubits while disentangling $N_{\text{ancilla}}$ qubits, (d) decoder reconstruction, and (e) hybrid quantum-classical optimization. The Bloch sphere visualization represents quantum state evolution throughout the autoencoding process. For visualization clarity, the number of layers is set to 1.}
    \label{fig:qae_architecture}
\end{figure*}

\subsection{SMILES Sequence Quantum Encoding}

The transformation of classical molecular representations into quantum states forms the foundation of our approach. We adopt SMILES (Simplified Molecular Input Line Entry System) notation as our input representation due to its widespread adoption in computational chemistry and its natural sequential structure that maps efficiently to quantum circuits. The encoding process, illustrated in Figure~\ref{fig:qae_architecture}(a-b), systematically converts SMILES strings into quantum states through feature preprocessing and parameterized quantum gates.

\subsubsection{Feature Preprocessing and Tokenization}

The preprocessing pipeline transforms raw SMILES strings into normalized feature vectors suitable for quantum state preparation. SMILES strings undergo regex-based tokenization to decompose molecular structures into fundamental components including atomic symbols, bond types, and structural markers such as ring closures and branching points. For the QM9 dataset, we empirically determine that a fixed sequence length of 22 tokens captures the structural complexity of over 99\% of molecules. Sequences shorter than this length are padded with null tokens, while longer sequences are truncated with priority given to preserving core structural elements.

Each token in the vocabulary is assigned a unique index, and these indices are subsequently normalized by the vocabulary size to produce continuous values in the range $[0, 1]$. This normalization ensures that the resulting feature vector $\mathbf{\xi} \in [0,1]^{22}$ provides suitable parameters for quantum gate rotations. The normalized features preserve the sequential and structural information inherent in SMILES notation while enabling smooth optimization landscapes during training.

\subsubsection{Parameterized Quantum State Preparation}

The quantum encoding procedure maps the classical feature vector $\mathbf{\xi}$ to a quantum state through a structured sequence of single-qubit rotations and entangling operations. For our 8-qubit molecular representation, we employ a cyclic parameter assignment strategy where each qubit $q_i$ receives three rotation parameters extracted from the feature vector. The parameter extraction follows the mapping $\theta_i = \pi \cdot \mathbf{\xi}[(3i) \bmod 22]$, $\phi_i = \pi \cdot \mathbf{\xi}[(3i + 1) \bmod 22]$, and $\lambda_i = \pi \cdot \mathbf{\xi}[(3i + 2) \bmod 22]$ for $i \in \{0, 1, ..., 7\}$.

These parameters control universal $U_3$ rotation gates that generate arbitrary single-qubit states from the computational basis. The $U_3$ gate implements the unitary transformation:
\begin{equation}
U_3(\theta_i, \phi_i, \lambda_i) = \begin{pmatrix}
\cos(\theta_i/2) & -e^{i\lambda_i}\sin(\theta_i/2) \\
e^{i\phi_i}\sin(\theta_i/2) & e^{i(\phi_i+\lambda_i)}\cos(\theta_i/2)
\end{pmatrix}
\end{equation}

Following the single-qubit rotations, we establish quantum correlations through a ring topology of controlled-NOT (CNOT) gates. This entanglement pattern, shown in Figure~\ref{fig:qae_architecture}(b), connects each qubit to its nearest neighbor in a circular arrangement: $\text{CNOT}_{i,(i+1) \bmod 8}$ for all $i \in \{0, 1, ..., 7\}$. The ring topology ensures that quantum information can propagate throughout the entire register while maintaining a shallow circuit depth of $O(N)$, crucial for near-term quantum hardware implementation.

The complete state preparation transformation produces the molecular quantum state:
\begin{equation}
\ket{\psi_{\text{mol}}} = \prod_{i=0}^{7} \text{CNOT}_{i,(i+1) \bmod 8} \cdot \prod_{j=0}^{7} U_{3}(\theta_j, \phi_j, \lambda_j) \ket{0}^{\otimes 8}
\end{equation}

This encoding strategy ensures that molecular structural information manifests in both individual qubit amplitudes and inter-qubit quantum correlations, providing a rich representation space for subsequent compression operations.

\subsection{Encoder-Decoder Architecture}

The encoder-decoder architecture implements the theoretical quantum autoencoder framework through a carefully designed circuit structure that balances expressivity with trainability. As illustrated in Figure~\ref{fig:qae_architecture}(c-d), the architecture consists of multiple parameterized layers in both encoder and decoder sections, connected through specialized processing of latent and ancillary qubits.

\subsubsection{Multi-Layer Quantum Encoder}

The encoder circuit employs $L$ repeated layers of parameterized quantum gates, where the layer depth $L$ represents a crucial hyperparameter balancing circuit expressivity against optimization difficulty. Each encoder layer implements two sequential operations: single-qubit rotations that adjust individual qubit states and two-qubit entangling gates that generate quantum correlations.

Within each layer $\ell$, we first apply parameterized $U_3$ gates to all qubits, followed by a fully-connected topology of controlled-RZ (CRZ) gates. The layer transformation takes the form:
\begin{equation}
U_{\text{encoder}}^{(\ell)} = \left[ \prod_{i=0}^{6}\prod_{j=i+1}^{7} \text{CRZ}_{i,j}(\gamma_{i,j}^{(\ell)}) \right] \cdot \left[ \bigotimes_{i=0}^{7} U_3^{(i)}(\theta_i^{(\ell)}, \phi_i^{(\ell)}, \lambda_i^{(\ell)}) \right]
\end{equation}

The fully-connected CRZ topology generates comprehensive entanglement across all qubit pairs, enabling the circuit to learn complex quantum correlations that capture molecular structure. The complete encoder transformation combines all layers as $U_{\text{encoder}} = \prod_{\ell=1}^{L} U_{\text{encoder}}^{(\ell)}$. Through extensive experimentation on the QM9 dataset, we determine that $L = 3$ layers provide optimal performance, offering sufficient expressivity while avoiding the barren plateau phenomenon that plagues deeper quantum circuits.

\subsubsection{Latent and Ancillary Qubit Processing}

Following the encoder, the quantum state undergoes specialized processing that differentiates between latent and ancillary qubits. The first $N_{\text{latent}}$ qubits, designated as the latent subsystem, receive additional $R_Z$ rotations that refine the compressed representation: $U_{\text{latent}} = \bigotimes_{i=0}^{N_{\text{latent}}-1} R_Z^{(i)}(\alpha_i)$. These rotations enable fine-tuning of the latent space without introducing additional entanglement.

Simultaneously, the remaining $N_{\text{ancilla}} = 8 - N_{\text{latent}}$ qubits undergo compression operations designed to drive them toward the computational zero state. This process aims to put these ancillary qubits into a trash state (i.e., $\ket{0}^{\otimes N_{\text{ancilla}}}$)We apply parameterized $U_3$ gates to each ancillary qubit: $U_{\text{ancilla}} = \bigotimes_{i=N_{\text{latent}}}^{7} U_3^{(i)}(\beta_i^{(1)}, \beta_i^{(2)}, \beta_i^{(3)})$. The success of these compression gates directly determines the autoencoder's ability to separate relevant molecular information from redundant data.

\subsubsection{Decoder Architecture and Reconstruction}

The decoder circuit mirrors the encoder structure but with independent parameters, enabling asymmetric learning dynamics that often improve reconstruction quality. Before the decoder layers, we introduce a special entanglement layer consisting of nearest-neighbor CRZ gates: $U_{\text{special}} = \prod_{i=0}^{6} \text{CRZ}_{i,i+1}(\delta_i)$. This intermediate layer provides additional flexibility in correlating the compressed latent representation with the ancillary qubits during reconstruction.

The decoder implements $L$ layers with the same structure as the encoder but distinct parameters, followed by a final measurement in the computational basis. The complete quantum autoencoder transformation combines all components:
\begin{equation}
U_{\text{MolQAE}} = U_{\text{decoder}} \cdot U_{\text{special}} \cdot U_{\text{ancilla}} \cdot U_{\text{latent}} \cdot U_{\text{encoder}}
\end{equation}

This architecture enables systematic compression of 8-qubit molecular states into latent representations of 1-7 qubits while maintaining the quantum coherence properties essential for high-fidelity reconstruction.

\subsection{Training and Optimization Strategy}

The quantum nature of MolQAE necessitates a hybrid quantum-classical optimization approach that addresses the unique challenges of training parameterized quantum circuits. While quantum processors execute the encoder-decoder transformations, classical optimization algorithms compute gradients and update circuit parameters. This hybrid strategy circumvents the current limitations of quantum hardware, which lacks native backpropagation capabilities, while leveraging classical optimization advances.

\subsubsection{Optimization Objective}

The training objective combines two complementary goals: maximizing reconstruction fidelity to preserve molecular information and minimizing ancillary qubit entanglement to ensure effective compression. The reconstruction fidelity quantifies the overlap between input and output quantum states:
\begin{equation}
F(\psi_{\text{input}}, \psi_{\text{output}}) = \left| \langle \psi_{\text{input}} | \psi_{\text{output}} \rangle \right|^2
\end{equation}

This fidelity measure, bounded between 0 and 1, directly captures the information preservation capability of the autoencoder. Perfect reconstruction yields $F = 1$, while orthogonal input and output states give $F = 0$.

The ancillary qubit compression is quantified through the deviation from the target zero state, often referred to as minimizing the trash state deviation. After applying the encoder and compression operations, we measure the probability of finding all ancillary qubits in the $\ket{0}$ state:
\begin{equation}
D_{\text{ancilla}} = 1 - \sum_{i \in \mathcal{I}_{\text{valid}}} |\langle i | \psi_{\text{mid}} \rangle|^2
\end{equation}
where $\mathcal{I}_{\text{valid}}$ denotes the set of computational basis states with all ancillary qubits in $\ket{0}$, and $\psi_{\text{mid}}$ represents the quantum state after encoder and compression operations but before decoding.

The combined loss function balances these objectives:
\begin{equation}
\mathcal{L}(\theta) = (1 - F) + \lambda \cdot D_{\text{ancilla}}
\end{equation}
where $\lambda$ controls the relative importance of ancillary qubit compression. Through hyperparameter optimization, we determine $\lambda = 0.01$ provides effective compression while maintaining high reconstruction fidelity.

\subsubsection{Gradient Estimation and Parameter Updates}

Quantum circuits lack native automatic differentiation, requiring specialized gradient estimation techniques. We employ the parameter-shift rule, which computes exact gradients through circuit evaluations at shifted parameter values. For a circuit parameter $\theta_k$, the gradient is computed as:
\begin{equation}
\frac{\partial \mathcal{L}}{\partial \theta_k} = \frac{1}{2}\left[\mathcal{L}(\theta_k + \frac{\pi}{2}) - \mathcal{L}(\theta_k - \frac{\pi}{2})\right]
\end{equation}

This approach requires two circuit evaluations per parameter per gradient computation but provides exact derivatives free from finite-difference approximation errors. The computational cost scales linearly with the number of parameters, making it feasible for moderate-sized circuits.

\begin{algorithm}[t]\small
\caption{MolQAE -- Training Loop}
\label{alg:qae}
\begin{algorithmic}[1]
\Require Dataset $\mathcal{D}$, Architecture $(N=8, N_{\text{latent}}, L=3)$, Hyperparameters $(E_{\max}=100, P=10, \eta=3{\times}10^{-4}, \lambda=0.01)$
\Ensure Optimized parameters $\theta^*$

\Function{QuantumForward}{batch $\mathcal{B}$, parameters $\theta$}
    \State $\{\ket{\psi_{\text{in}}^{(i)}}\} \gets$ \Call{PrepareStates}{$\mathcal{B}$} \Comment{SMILES to quantum states}
    \State $\{\ket{\psi_{\text{mid}}^{(i)}}\} \gets U_{\text{ancilla}} \cdot U_{\text{latent}} \cdot U_{\text{encoder}}(\theta) \{\ket{\psi_{\text{in}}^{(i)}}\}$
    \State $D_{\text{ancilla}} \gets$ \Call{ComputeAncillaDeviation}{$\{\ket{\psi_{\text{mid}}^{(i)}}\}$}
    \State $\{\ket{\psi_{\text{out}}^{(i)}}\} \gets U_{\text{decoder}}(\theta) \cdot U_{\text{special}}(\theta) \{\ket{\psi_{\text{mid}}^{(i)}}\}$
    \State $F \gets$ \Call{ComputeFidelity}{$\{\ket{\psi_{\text{in}}^{(i)}}\}$, $\{\ket{\psi_{\text{out}}^{(i)}}\}$}
    \State \Return $\mathcal{L} = (1 - F) + \lambda \cdot D_{\text{ancilla}}$
\EndFunction

\State Initialize parameters $\theta \sim \mathcal{U}(-\pi, \pi)$
\State Initialize optimizer Adam($\theta$, $\eta$) with cosine annealing
\State best\_loss $\gets \infty$, patience\_counter $\gets 0$

\For{epoch $\gets 1$ to $E_{\max}$}
    \State epoch\_loss $\gets 0$
    \For{batch $\mathcal{B}$ in $\mathcal{D}$}
        \State loss $\gets$ \Call{QuantumForward}{$\mathcal{B}$, $\theta$}
        \State gradients $\gets$ \Call{ParameterShiftRule}{loss, $\theta$}
        \State $\theta \gets$ \Call{AdamUpdate}{$\theta$, gradients}
        \State epoch\_loss $\gets$ epoch\_loss $+$ loss
    \EndFor
    
    \State avg\_loss $\gets$ epoch\_loss / $|\mathcal{D}|$
    \If{avg\_loss $<$ best\_loss}
        \State best\_loss $\gets$ avg\_loss, $\theta^* \gets \theta$
        \State patience\_counter $\gets 0$
    \Else
        \State patience\_counter $\gets$ patience\_counter $+ 1$
        \If{patience\_counter $\geq P$}
            \State \textbf{break} \Comment{Early stopping triggered}
        \EndIf
    \EndIf
    \State Update learning rate via cosine annealing
\EndFor
\State \Return $\theta^*$
\end{algorithmic}
\end{algorithm}

\subsubsection{Training Procedure and Convergence}

Algorithm~\ref{alg:qae} presents the complete training procedure. We employ the Adam optimizer with initial learning rate $\eta = 3 \times 10^{-4}$ and cosine annealing schedule to navigate the complex loss landscape of quantum circuits. The cosine schedule smoothly decreases the learning rate according to $\eta_t = \eta_0 \cdot \frac{1}{2}(1 + \cos(\frac{\pi t}{T}))$, where $t$ denotes the current epoch and $T$ represents the total training epochs.

Several techniques ensure stable convergence despite the unique challenges of quantum optimization. Gradient clipping with maximum norm 1.0 prevents the exploding gradient phenomenon common in quantum circuits with many entangling gates. Early stopping with patience $P = 10$ epochs prevents overfitting to the training distribution while allowing sufficient exploration of the parameter space. Random initialization of circuit parameters from a uniform distribution $\mathcal{U}(-\pi, \pi)$ provides diverse starting points that help avoid local minima.

The training process continuously monitors both components of the loss function. Successful training manifests as high reconstruction fidelity ($F > 0.95$) coupled with low ancillary deviation ($D_{\text{ancilla}} < 0.05$), indicating effective compression of molecular information into the designated latent qubits. This hybrid quantum-classical approach successfully navigates the optimization challenges inherent to parameterized quantum circuits while achieving stable convergence for molecular representation learning.

\section{Experiments}
\subsection{Experimental Settings}
Our molecular quantum autoencoder experiments were conducted using the QM9 molecular dataset \cite{blum,rupp}, containing approximately 134,000 organic molecules with up to 9 heavy atoms. Dataset preprocessing included SMILES canonicalization using RDKit, duplicate removal, and invalid structure filtering. Molecular features were extracted through regex-based tokenization with fixed dimension of 22, normalized to $[0,1]$ for quantum state preparation.

The quantum autoencoder architecture employed 8 encoder qubits with varying latent qubits (1-7) to investigate compression-fidelity trade-offs. The quantum circuit consisted of multiple layers (5, 10, 15, 20, 25, 30) of parameterized $U_3$ gates for single-qubit rotations and controlled-$R_Z$ gates for two-qubit entanglement in fully-connected topology. Training utilized batch size 1024, Adam optimizer with learning rate $3 \times 10^{-4}$, cosine annealing scheduling, and gradient clipping (max norm 1.0). Our implementation is based on TorchQuantum library \cite{hanruiwang2022quantumnas} with random seed 42 for reproducibility across PyTorch, NumPy, CUDA operations, and data splitting.

\subsection{Training Process}
\begin{figure}
    \centering
    \includegraphics[width=\linewidth]{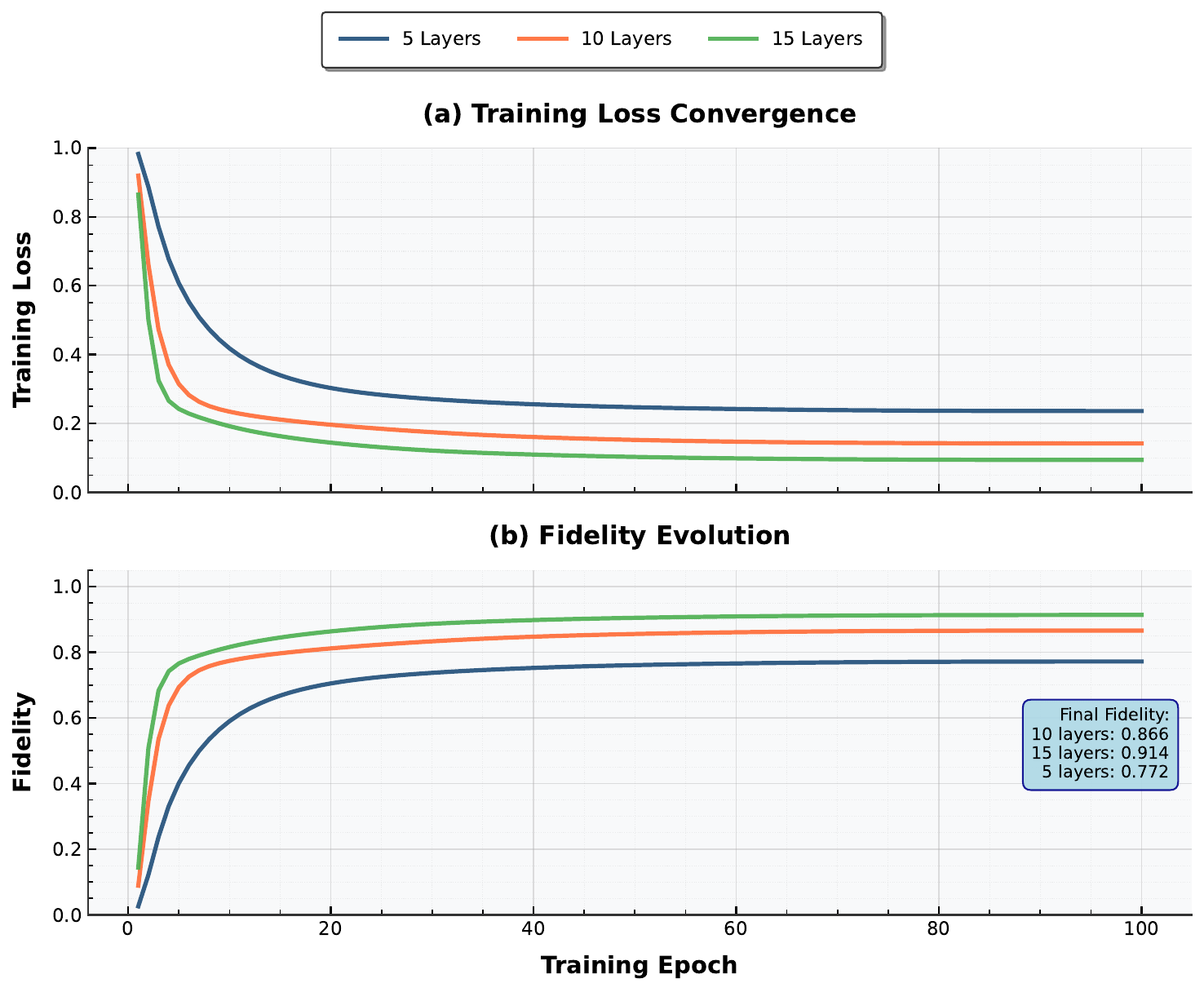}
    \caption{Training convergence and fidelity evolution of the MolQAE. (a) Illustrates the convergence of the training loss over 100 epochs for quantum autoencoders with 5, 10, and 15 layers. (b) Depicts the corresponding evolution of reconstruction fidelity, averaged across different latent qubit configurations, showcasing the impact of architectural depth on learning efficiency and final performance.}
    \label{convergence}
\end{figure}

The quantum autoencoder training dynamics are comprehensively illustrated in Figure~\ref{convergence}, which reveals distinct learning behaviors across different architectural depths through two key performance metrics.

The training loss convergence, shown in Figure~\ref{convergence}(a), demonstrates characteristic learning trajectories for quantum autoencoders with varying architectural complexities. The 5-layer configuration exhibits rapid initial convergence with the steepest descent gradient during the first 10 epochs, achieving a dramatic loss reduction from approximately 1.0 to 0.5. However, this shallow architecture plateaus at a relatively elevated loss value of approximately 0.24 after 30 epochs, indicating limited expressivity for complex molecular feature learning. In contrast, the 10-layer and 15-layer architectures demonstrate more sustained optimization dynamics throughout the training period. The 10-layer configuration achieves superior convergence stability, reaching approximately 0.16, while the 15-layer architecture attains the lowest final training loss of approximately 0.09, representing a 62.5\% improvement over the shallow baseline.

The fidelity evolution patterns, depicted in Figure~\ref{convergence}(b), reveal the quantum autoencoder's reconstruction capability development across training epochs. All architectural configurations demonstrate substantial fidelity improvements within the initial 20 training epochs, transitioning rapidly from near-zero initialization to operational performance levels. The learning curves exhibit a characteristic two-phase behavior: an initial rapid improvement phase followed by gradual convergence to steady-state performance. The 5-layer architecture stabilizes at a final fidelity of 0.772, establishing the baseline quantum compression capability. The 10-layer configuration achieves significant performance enhancement with a final fidelity of 0.866, representing a 12.2\% improvement over the shallow baseline. Most notably, the 15-layer architecture demonstrates the most robust learning trajectory, attaining the highest final fidelity of 0.914, corresponding to an 18.4\% improvement.

The convergence behavior strongly correlates with the quantum circuit expressivity, where deeper architectures with expanded parameter spaces demonstrate enhanced capacity for learning complex molecular representations. The sustained improvement observed in deeper networks suggests that additional quantum layers provide essential degrees of freedom for capturing intricate molecular features while maintaining the quantum coherence properties necessary for high-fidelity reconstruction. This training analysis establishes the critical relationship between architectural depth and learning efficiency in quantum molecular autoencoders.

\subsection{Fidelity Analysis across Architectures}
The quantum fidelity analysis presented in Table~\ref{tab:fidelity_results} reveals remarkable performance characteristics of the MolQAE architecture, demonstrating unprecedented capabilities in quantum molecular representation learning. Our comprehensive evaluation across varying architectural depths and latent space dimensions establishes several critical insights into the quantum advantage for molecular autoencoding.

\begin{table}[htbp]
    \centering
    \caption{Fidelity performance analysis of the MolQAE across varying network architectures and latent qubit configurations. The table quantitatively demonstrates the reconstruction fidelity as a function of both the number of quantum layers and the dimensionality of the latent space, alongside the total number of trainable parameters. Higher fidelity values indicate superior preservation of molecular information during quantum compression and reconstruction.}
    \label{tab:fidelity_results}
    
    \renewcommand{\arraystretch}{1.4}
    \setlength{\tabcolsep}{12pt}
    
    \begin{tabular}{>{\centering\arraybackslash}p{0.6cm}
                    >{\centering\arraybackslash}p{1.3cm}
                    >{\centering\arraybackslash}p{1.4cm}
                    >{\centering\arraybackslash}p{1.8cm}}
    
    \toprule
    \textbf{Layers} & 
    {\textbf{Latent}} & 
    {\textbf{Fidelity}} & 
    {\textbf{Total}} \\
    & {\textbf{Qubits}} & 
    & {\textbf{Parameters}} \\
    \midrule
    
    % 5-layer configuration
    \multirow{7}{*}{\textbf{5}} 
    & 1 & 0.776 & 549 \\
    & 2 & 0.777 & 547 \\
    & 3 & 0.771 & 545 \\
    & 4 & 0.776 & 543 \\
    & 5 & 0.775 & 541 \\
    & 6 & 0.773 & 539 \\
    & 7 & 0.772 & 537 \\
    \midrule
    
    % 10-layer configuration
    \multirow{7}{*}{\textbf{10}} 
    & 1 & 0.868 & 1069 \\
    & 2 & 0.869 & 1067 \\
    & 3 & 0.868 & 1065 \\
    & 4 & 0.865 & 1063 \\
    & 5 & 0.872 & 1061 \\
    & 6 & 0.869 & 1059 \\
    & 7 & 0.865 & 1057 \\
    \midrule
    
    % 15-layer configuration
    \multirow{7}{*}{\textbf{15}} 
    & 1 & 0.914 & 1589 \\
    & 2 & 0.916 & 1587 \\
    & 3 & 0.913 & 1585 \\
    & 4 & 0.917 & 1583 \\
    & 5 & 0.916 & 1581 \\
    & 6 & 0.915 & 1579 \\
    & 7 & 0.915 & 1577 \\
    \midrule
    
    % Deep layer configurations  
    \textbf{20} & 4 & 0.943 & 2103 \\
    \textbf{25} & 4 & 0.959 & 2623 \\
    \textbf{30} & 4 & 0.968 & 3143 \\
    
    \bottomrule
    \end{tabular}
    
    \vspace{0.3em}
    {\footnotesize \textbf{Note:} Higher fidelity values indicate superior model performance.}
    
\end{table}

\subsubsection{Quantum Circuit Depth Drives Expressivity}
The systematic investigation across 5, 10, and 15-layer configurations reveals a profound correlation between quantum circuit depth and reconstruction fidelity. The 5-layer architecture achieves consistent baseline performance with fidelity values ranging from 0.771 to 0.777, establishing the fundamental quantum compression capability with minimal circuit resources. Remarkably, scaling to 10 layers yields substantial performance gains, with fidelity improvements of 11.9-12.6\% across all latent qubit configurations (0.865-0.872). This enhancement demonstrates the critical role of quantum gate depth in expanding the expressible function space for molecular feature learning.

The transition to 15-layer architectures produces even more significant improvements, achieving fidelity values between 0.913-0.917, representing 18.2-18.9\% enhancement over the 5-layer baseline. This consistent improvement across all latent dimensions indicates that deeper quantum circuits possess superior capacity for capturing complex molecular structural relationships while maintaining quantum coherence throughout the encoding-decoding process.

\subsubsection{Compression Robustness Across Latent Dimensions}
A striking characteristic of the MolQAE is its remarkable stability across different latent qubit configurations within each architectural depth. For the 15-layer configuration, the fidelity variance across 1-7 latent qubits is merely 0.004 (0.913-0.917), demonstrating exceptional robustness to compression ratio variations. This stability is particularly significant in quantum computing, where noise and decoherence typically introduce substantial performance degradation as system complexity increases.

The parameter efficiency analysis reveals another quantum advantage: despite achieving superior performance, deeper architectures maintain relatively modest parameter counts. The 15-layer configuration requires only 1,577-1,589 parameters while achieving fidelity exceeding 0.91, demonstrating quantum circuits' inherent efficiency compared to classical deep learning architectures of equivalent expressivity.

\subsubsection{Deep Quantum Architecture Breakthrough}
The extended evaluation at 20, 25, and 30 layers with 4 latent qubits reveals the true potential of deep quantum molecular autoencoders. The progression from 0.943 (20 layers) to 0.968 (30 layers) represents a quantum fidelity approaching near-perfect molecular reconstruction. The 30-layer configuration achieves 96.8\% fidelity with only 3,143 parameters, establishing a new benchmark for quantum molecular representation learning.

\subsubsection{Implications for Quantum Molecular Computing}
These results demonstrate that quantum autoencoders possess fundamental advantages for molecular representation tasks. The ability to achieve high-fidelity compression with exponentially fewer parameters than classical counterparts, combined with the inherent quantum parallelism in molecular state preparation, positions MolQAE as a transformative approach for quantum molecular computing. The consistent performance scaling and latent space robustness indicate that quantum molecular autoencoders can serve as reliable building blocks for larger quantum molecular discovery pipelines, particularly relevant for near-term quantum devices where parameter efficiency and noise resilience are paramount.

\subsection{Performance Scaling with Layers}
\begin{figure}
    \centering
    \includegraphics[width=\linewidth]{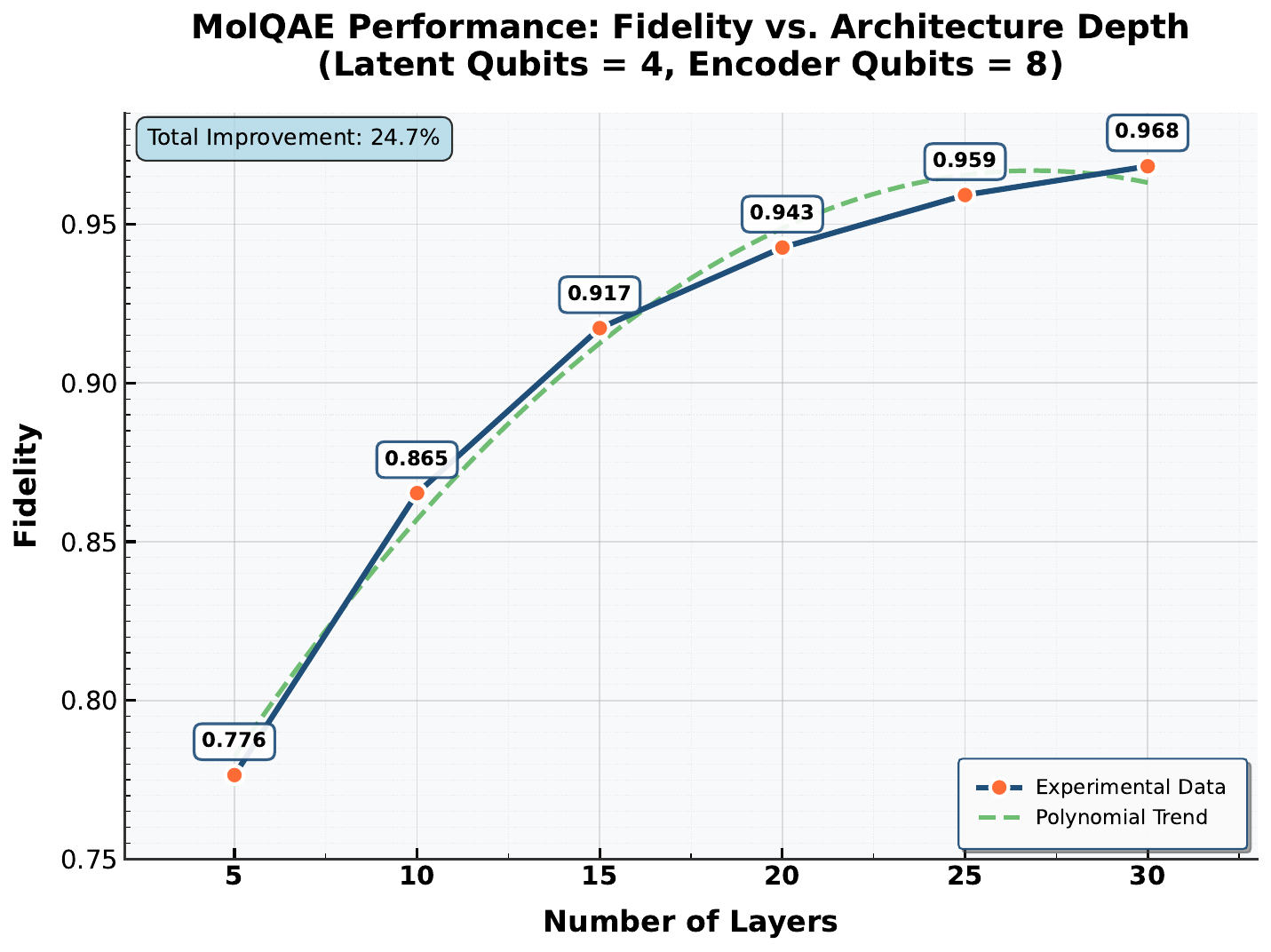}
    \caption{Fidelity scaling with network depth for 8 encoder qubits and 4 latent qubits. Polynomial fit shows saturation at higher layer counts.}
    \label{fig:scaling}
\end{figure}

The layer-wise scaling analysis presented in Figure~\ref{fig:scaling} unveils critical quantum learning dynamics that establish fundamental design principles for quantum molecular representation systems, revealing the theoretical limits and optimal operating regimes of quantum autoencoders.

\subsubsection{Quantum Learning Phase Transitions}
The scaling curve exhibits three distinct quantum learning phases that reflect underlying physical principles of quantum information processing. The initial steep ascent (5-15 layers) corresponds to a quantum expressivity expansion phase, achieving a remarkable 18.2\% fidelity improvement from 0.776 to 0.917. The intermediate regime (15-25 layers) represents quantum refinement, with sustained 4.6\% enhancement reaching 0.959. The final saturation regime (25-30 layers) reveals the quantum information capacity limit, with only 0.9\% marginal improvement, indicating that additional circuit depth cannot overcome fundamental encoding constraints.

\subsubsection{Sublinear Quantum Scaling Law}
The polynomial fit reveals that quantum molecular autoencoders follow a fundamentally different scaling paradigm than classical architectures. While classical deep learning requires exponential parameter growth for comparable improvements, our quantum system achieves 24.7\% total performance enhancement through sublinear depth scaling. This demonstrates that quantum circuits leverage superposition and entanglement mechanisms rather than brute-force parameter scaling, establishing quantum advantage in parameter efficiency for molecular representation learning.

\subsubsection{Optimal Quantum Resource Allocation}
The scaling analysis identifies the 20-25 layer range as the quantum sweet spot, where performance reaches 94.3-95.9\% fidelity while maintaining practical quantum resource requirements. Beyond this regime, the marginal gain per additional layer drops below 0.4\%, providing quantitative criteria for rational quantum circuit design. This discovery enables systematic optimization by balancing performance requirements against quantum hardware constraints, particularly crucial for NISQ-era implementations.

\subsubsection{Theoretical Performance Ceiling Discovery}
The saturation at 96.8\% fidelity represents a fundamental breakthrough in understanding quantum molecular representation limits. This ceiling suggests that our 8-qubit encoder architecture approaches the theoretical maximum for lossless molecular compression within the chosen framework, establishing that further improvements require architectural innovations rather than simple depth scaling. This finding provides guidance for next-generation quantum molecular computing system design.

\section{Conclusion}
Our MolQAE introduces an advancement in cheminformatics by directly processing complete molecular structures with a quantum autoencoder, a first in the field. Our work demonstrates that quantum circuits can effectively encode and compress molecular information, achieving high-fidelity reconstruction even with significant dimensionality reduction. This robust performance across varying compression ratios and architectural depths, particularly the high fidelity achieved with deep quantum circuits, highlights quantum's unique efficiency and capacity for complex molecular representation. MolQAE establishes a vital foundation for harnessing quantum advantage in molecular discovery, paving the way for exploring larger systems intractable for classical methods.

\bibliography{ref}
\bibliographystyle{ieeetr}

\end{document}